\documentclass{elsart}

\usepackage{graphicx}
\usepackage{booktabs}
\usepackage{amsmath, amssymb}
\usepackage{hyperref}
\usepackage[square,comma]{natbib}
\usepackage{pxfonts}
\usepackage{lineno}

\journal{}

\usepackage{lineno}
\begin{document}

\thispagestyle{empty}
\begin{Large}
\textbf{DEUTSCHES ELEKTRONEN-SYNCHROTRON}

\textbf{\large{Ein Forschungszentrum der
Helmholtz-Gemeinschaft}\\}
\end{Large}

DESY 17-054

March 2017

\begin{eqnarray}
\nonumber &&\cr \nonumber && \cr \nonumber &&\cr
\end{eqnarray}
\begin{eqnarray}
\nonumber
\end{eqnarray}
\begin{center}
\begin{Large}
\textbf{Automatic tuning of Free Electron Lasers}
\end{Large}

\begin{eqnarray}
\nonumber &&\cr \nonumber && \cr
\end{eqnarray}

\begin{large}
Ilya Agapov and Igor Zagorodnov
\end{large}
\textsl{\\Deutsches Elektronen-Synchrotron DESY, Hamburg, Germany}

\begin{large}
Gianluca Geloni,
\end{large}
\textsl{\\European XFEL, Schenefeld, Germany}

\begin{large}
Sergey Tomin
\end{large}
\textsl{\\European XFEL, Schenefeld, Germany}
\textsl{\\NRC "Kurchatov Institute", Moscow, Russia}

\begin{eqnarray}
\nonumber
\end{eqnarray}
\begin{eqnarray}
\nonumber
\end{eqnarray}
ISSN 0418-9833
\begin{eqnarray}
\nonumber
\end{eqnarray}
\begin{large}
\textbf{NOTKESTRASSE 85 - 22607 HAMBURG}
\end{large}

\end{center}
\clearpage
\newpage
\begin{frontmatter}

\title{Automatic tuning of Free Electron Lasers}

\author[DESY]{I. Agapov}
\author[XFEL]{G. Geloni}
\author[XFEL,NRC]{S. Tomin\thanksref{corr},}
\thanks[corr]{Corresponding Author. E-mail address: sergey.tomin@xfel.eu}
\author[DESY]{and I. Zagorodnov}

\address[DESY]{Deutsches Elektronen Synchrotron, Notkestrasse 85, 22607 Hamburg, Germany}
\address[XFEL]{European XFEL GmbH, Holzkoppel 4, 22869 Schenefeld, Germany}
\address[NRC]{NRC "Kurchatov Institute", Akademika Kurchatova pl. 1, 123182 Moscow, Russia}

\begin{abstract}

Existing FEL facilities often suffer  from stability issues: so electron orbit, transverse electron optics, electron bunch compression and other parameters have to be readjusted often to account for drifts in performance of various components. 
The tuning procedures typically employed in operation are often manual and lengthy. We have been developing a combination of model-free and model-based automatic tuning methods to meet the needs of present and upcoming XFEL facilities. 
Our approach has been implemented at FLASH \cite{flash} to achieve automatic SASE tuning using empirical control of orbit, electron optics and bunch compression. 
In this paper we describe our approach to empirical tuning, the software which implements it, and the results of using it at FLASH.
We also discuss the potential of using machine learning and model-based techniques in tuning methods.

\end{abstract}

\end{frontmatter}

\section{Introduction}

The ultimate performance of accelerator-based facilities can only be achieved by careful alignment and calibration of all subsystems. In practice, however, that could be done only  to a certain
extent due to operational constraints. Especially in linac-based FEL facilities many parameters remain uncertain and operators often resort to empirical tuning. The time required for such tuning can be substantially reduced 
by introducing automated tools. This is particularly important with increasing pressure on availability and performance of modern light sources.

The empirical methods have been gaining popularity for optimization of accelerator facilities. Accelerator optics tuning has been discussed in \cite{kek}, \cite{huang}. 
Optimization of trajectory has been implemented at FERMI \cite{fermi}. 

In present work we  describe our approach to automatic tuning of FEL facilities, and show that combination of simple empirical rules, 
statistical learning and flexible controls can lead to significant speedup of tuning procedures. The concept is presented in Section \ref{sec:2a} . The optimization results for FLASH are discussed in Section \ref{sec:flash} .

To facilitate creating of tuning strategies, an on-line model is provided with the software. The online model based on OCELOT \cite{ocelot} is briefly described in Section \ref{sec:2b}. 
However, in a machine like FLASH \cite{flash} where the accelerating fields are subject to jitter and collective effects play a significant role (see e.g. \cite{dohlus},\cite{scholz}),
it is difficult to have an accurate optics model. For example, only measured response matrices can be used for orbit manipulation.
So far, the model has found limited application for FLASH. It is expected that it could be more useful for the next generation of machines where the performance may be better understood due to more advanced diagnostics.

The possibility of developing a fully automated self-learning intelligent operator-replacing software employing machine learning methods can be advocated.
When trying to extend our software with more intelligent control logic by introducing accelerator physics concepts 
such as possibility of correcting orbit, optics etc. in an autonomous way we found that while making the whole system more complicated, not much was gained.
We pursue the physics-model-free approach and discuss the issue of coupling statistics and machine learning with the empirical tuning in Sections \ref{sec:2c} and \ref{sec:statistics}
on the examples of knob ranking.

The software for which this paper is a short introduction is freely distributed with the OCELOT framework \cite{ocelot} and has already found application at other facilities outside DESY (see \cite{optim2}, \cite{optim3}).
Its implementation is briefly described in Section \ref{sec:2d}. Both the software and the approach are completely general, they can be applied to a wide variety of optimization problems.
So, it has been used for empirical optimization of the injection into a storage ring which is briefly discussed in Section \ref{sec:siberia} .

Novel optimization methods based on the OCELOT framework are being investigated for LCSL \cite{optim3}. This topic is beyond the scope of the present paper. 
Empirical optimization software based on OCELOT will be included into the control system of European XFEL \cite{xfeltdr},
which is currently under commissioning at DESY.

\section{Approach}

\subsection{Tuning concept}\label{sec:2a}

The tuning is done by an agent (the optimizer). The agent performs a sequence of actions. An action is generally a maximization/minimization problem aimed at reaching certain goals related to measured beam parameters,
using a group of devices. The devices can be any set of magnets, RF parameters, or anything controllable. The goals can be photon pulse energy (also referred to as SASE level) measured with various types of detectors and subject
to averaging procedures (e.g. averaging over the bunch train), photon pulse spectrum, photon or electron orbit, or a combination of any of those. If an action fails to improve the objective, the parameters are rolled back and the next action starts.
In practice, the action is typically either a photon pulse energy maximization or photon beam positioning. Electron orbit can as well be controlled. However, this is rarely used since the optimization is typically done on the photon beam parameters 
directly.
For FLASH the radiation can cause demagnetization of undulators, and a beam loss penalty is always added. The beam loss penalty is proportional to the BLM signal when the losses are below 70\% of the BLM alarm level, 
proportional to the BLM signal with a larger proportionality coefficient when the BLM signal is between 70\% and 100\% of the alarm level, and is a large number when the alarm level is exceeded. With such penalty beam losses never occurred during our test
(unless the beam transmission was already poor initially, in which case the beam losses were eliminated by manual tuning first).  For more details concerning optimization experience at FLASH see Section \ref{sec:flash} .
The logic is compactly programmable as a python script. Each sequence is invoked by an operator who also decides on the stopping criteria of the whole optimization.
Such procedure roughly mimics what a human operator is doing, is simple and robust and forms the basis of our tuning software. Other tools such as the statistical analysis can provide additional input to the optimizer.
The concepts and first results for FLASH were initially presented in \cite{optim}.

\subsection{Online model}\label{sec:2b}

\begin{figure}[h!]
 \centering
 \includegraphics[width=80mm]{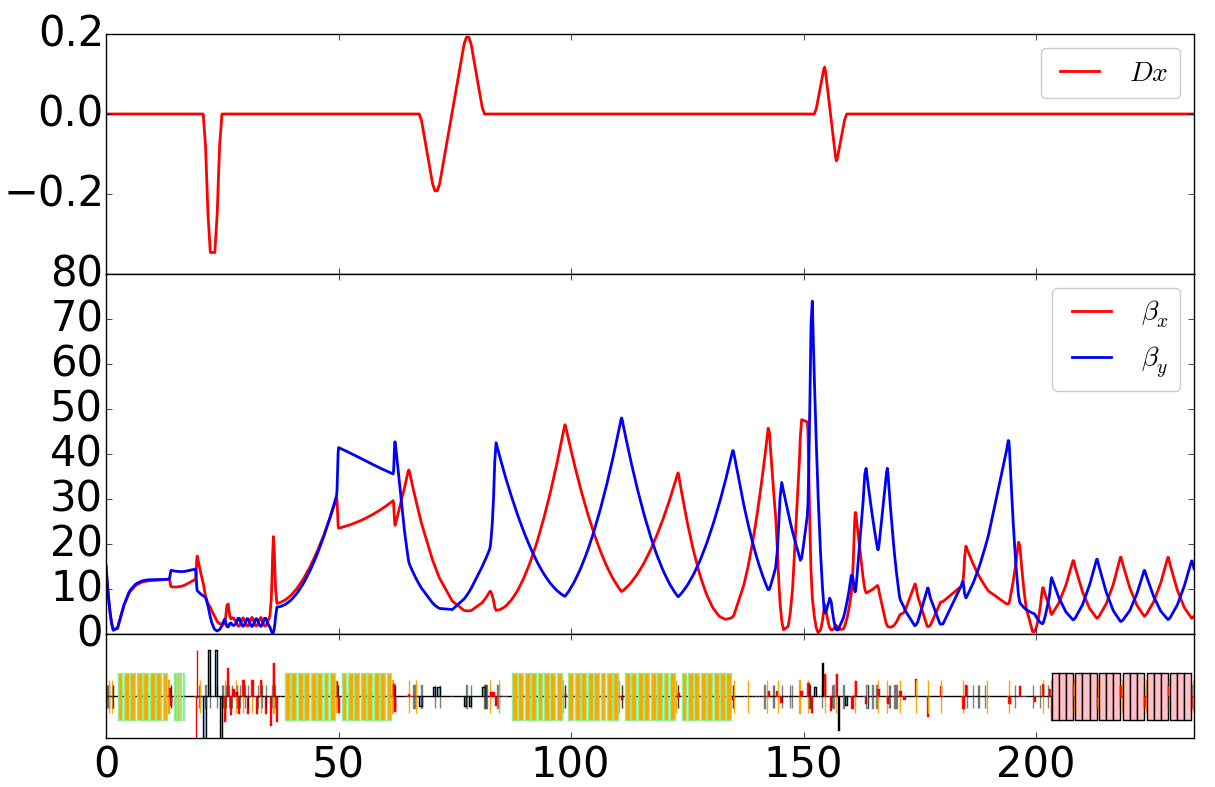}
 \caption{FLASH nominal optics in the online model}
 \label{fig:flash_nominal}
\end{figure}

\begin{figure}[h!]
 \centering
 \includegraphics[width=80mm]{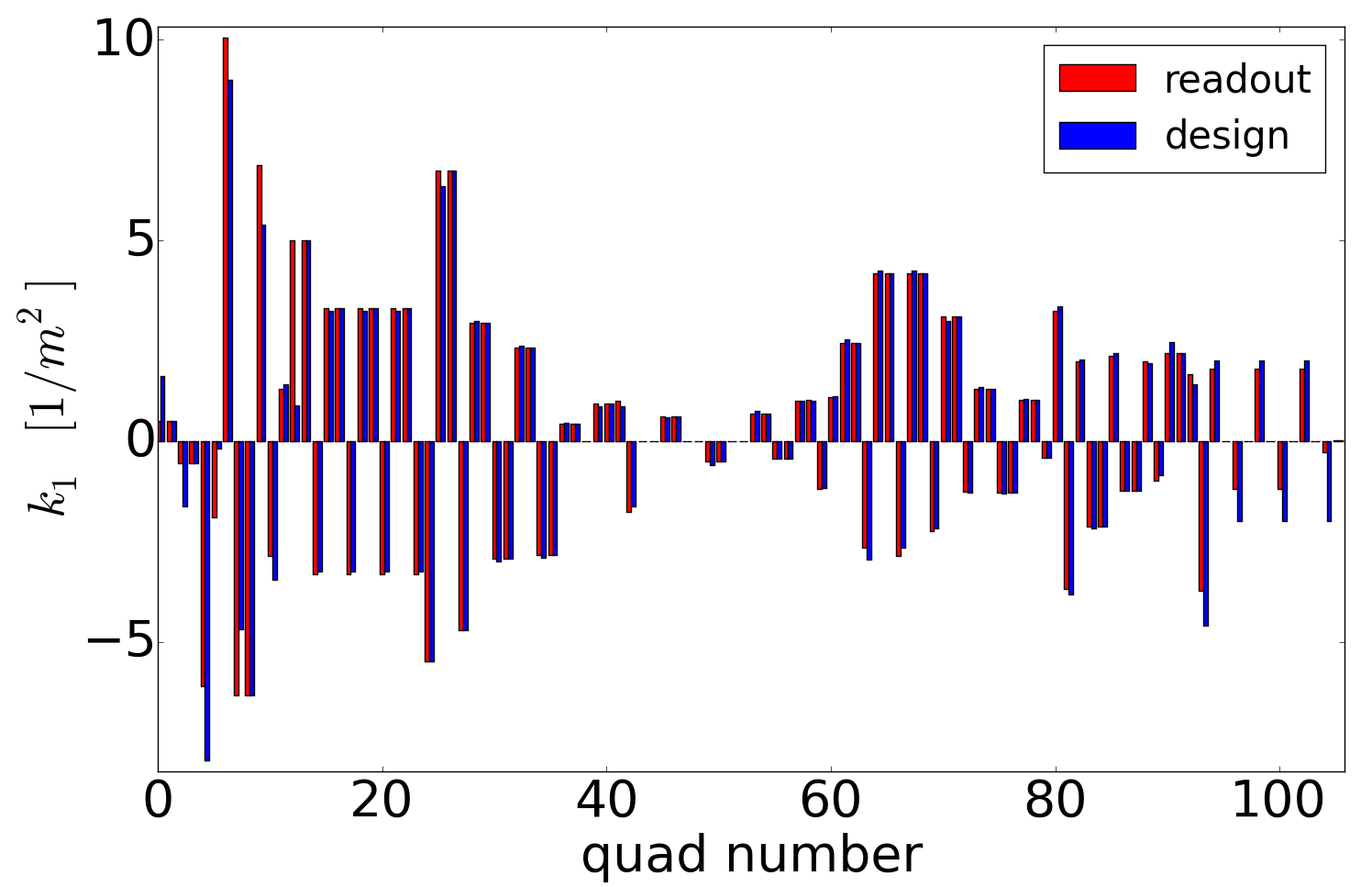}
 \caption{Comparison of theoretical quadrupole settings and those set in the machine}
 \label{fig:flash_kquad}
\end{figure}

One popular direction in high level control is on-line modeling, which typically involves creating a snapshot of current machine optics and performing various calculations on-line in the control room with such model.
The reason behind that is to speed up certain routine procedures where beam time is expensive and mitigate beam losses where these could be dangerous, for example in superconducting machines.
The capability of reading out magnet and RF settings from the machine and creating an OCELOT \cite{ocelot} optics model is also present in our toolkit. The initial reason for introducing such model was to facilitate ad-hoc multiknob creation such 
as closed bumps. This could allow creating more efficient tuning knobs by considering interdependencies between parameters. For example, SASE could be optimized with a closed bump through the linac while keeping
the orbit in the undulator section unaffected.
At FLASH, however, both theoretical and snapshot models are not accurate enough to allow creating such multiknobs for all machine sections. So, orbit manipulations require a measured response matrix. 
It is well known that the optics model cannot be uniquely inferred from a response matrix.
The procedure of deducing the model first (i.e. effectively measuring the optics after each change
of machine settings) thus appears too lengthy and ambiguous to be effectively integrated in the empirical tuning software, so the model is not made use of during the tuning.
This could be of more benefit for facilities such as the European XFEL \cite{xfeltdr} where
the amount of available hardware is substantially larger, there is more diagnostics, and such model could be used to guide a user to initially select the set of control parameters. 
Screenshots of online model for FLASH are shown in Figure \ref{fig:flash_nominal} for theoretical optics.
The difference between theoretical and snapshot settings is shown in Figure \ref{fig:flash_kquad}. The agreement for some parts of the machine is good. There the online model could successfully be used to perform beam
manipulations such as creating orbit bumps. However, for FLASH the difference proved to be too great to allow more interesting beam manipulations such as closed bumps through the linac.
While not of immediate use for the empirical tuning at FLASH, this model is developed further for European XFEL commissioning and operation. To this end, the relevant physics has been introduced into OCELOT, which includes space charge,
coherent synchrotron radiation, and wake fields; the details are discussed elsewhere \cite{ocelot2}. The idea is to use the model for beam manipulations in the machine sections where the model gives good agreement with reality, 
and purely empirical tuning where the agreement is poor. This is a direction of active research.

\subsection{Statistical analysis}\label{sec:2c}

A statistical learning problem is often reduced to deducing a functional dependency given a certain selection of input-output values \cite{stat1}.
We can consider the set of optimal controls (i.e. quadrupole and orbit corrector settings) as a deterministic function of a large set of parameters such as hardware temperatures, exact alignment, 
earth magnetic field and so forth. The nature of such dependency is typically hard to evaluate so we can consider such parameters effectively unknown, i.e. we can not deduce the controls with whatever measurements
available. While deducing the dependency of, say, the required orbit correction on the RF component temperatures might be hopeless, we can ask a) what kind of controls are best in a given situation and b) can we identify certain 
non-trivial dependencies in the data such as, for example, the 'golden orbit' as a function of charge. 
To address these questions for FLASH,  we can analyze three sources of data.
First, all accelerator parameters are constantly logged in an archive, which can in principle be analyzed. One could theoretically see what tuning sequences resulted in good performance.
However, the amount of data in this archive is enormous and extracting clean datasets is difficult. Just deducing which parameters were actually controlled and what time intervals are trustworthy in terms of diagnostics performance
is a non-trivial data analysis task. Even if such analysis were performed, we hardly expect new information than what can be learned from operator experience, i.e. that certain standard sets of devices were used and that the machine 
performance differed from run to run. We did not pursue this direction.

Second, the most important machine parameters are stored in a separate database on a regular basis, at least once per shift. This data is much more tangible and some analysis is presented in Section \ref{sec:statistics_a}.
While such analysis provides little guidance towards improving the tuning strategy, it gives useful information concerning both machine and the optimizer performance.

The problem with the latter data is that one could analyze the correlation between certain values but not the actions which led to the result. 
To improve this a third database was introduced to the optimizer which stores the machine state before and after each action as well as the action parameters such as the number of iterations of the optimization method. 
One could introduce figures of merit such as 

\begin{equation}\nonumber
 FoM_1 = S_2 - S_1 ; \quad FoM_2 = \frac{S_2 - S_1}{S_1}
\end{equation}

where $S_1$ and $S_2$ are initial and final pulse energies. Such performance criteria can be used to rank the optimization knobs, which is further discussed in Section \ref{sec:statistics_b}.

Clustering analysis of orbit data for FLASH revealed the dependence of the 'golden orbit' on the electron bunch charge.  Since our approach uses the direct photon pulse parameter optimization, 
such information could not yet be exploited and is not discussed here.

\subsection{Controls software}\label{sec:2d}

\begin{figure}[h!]
 \centering
 \includegraphics[width=80mm]{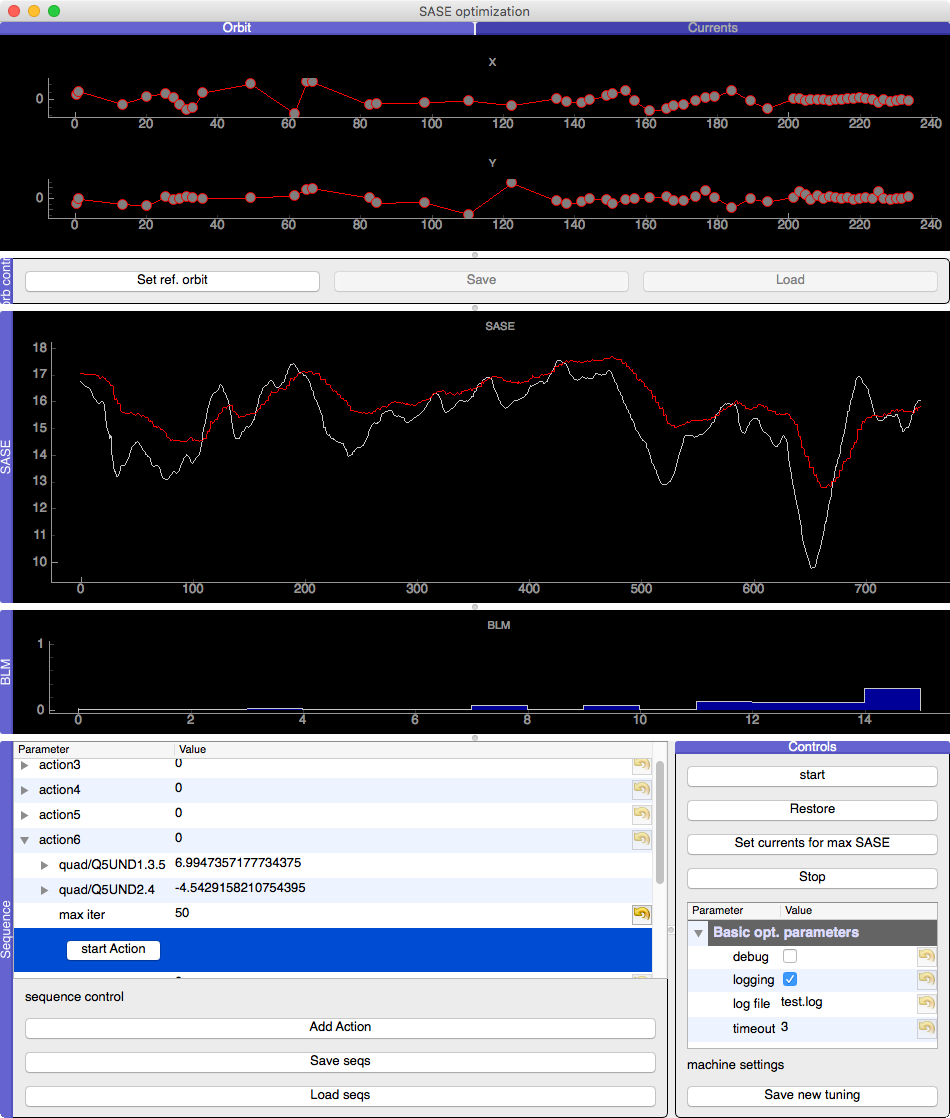}
 \caption{Optimizer GUI}
 \label{fig:gui}
\end{figure}

The optimization software is implemented in python with  DOOCS \cite{doocs} is used as a controls interface, and is freely distributed with the OCELOT framework. 
It can be run either in scripting mode with full control over the functionality exposed by the optimization module, or in a more operator-friendly graphical mode. A screenshot of the graphical interface 
is shown in Figure \ref{fig:gui}. The graphical mode allows to group devices into actions, set various run parameters, start and stop the optimization. The machine learning and statistical analysis features
make use of the {\it scikit-learn} \cite{sklearn} machine learning library. The numerical methods used in maximization/miminization can be adjusted and extended. Any functional minimization available with the
{\it scipy} \cite{scipy} package can be easily added.

\section{Empirical tuning at FLASH}\label{sec:flash}

\begin{figure}[h!]
 \centering
 \includegraphics[width=80mm]{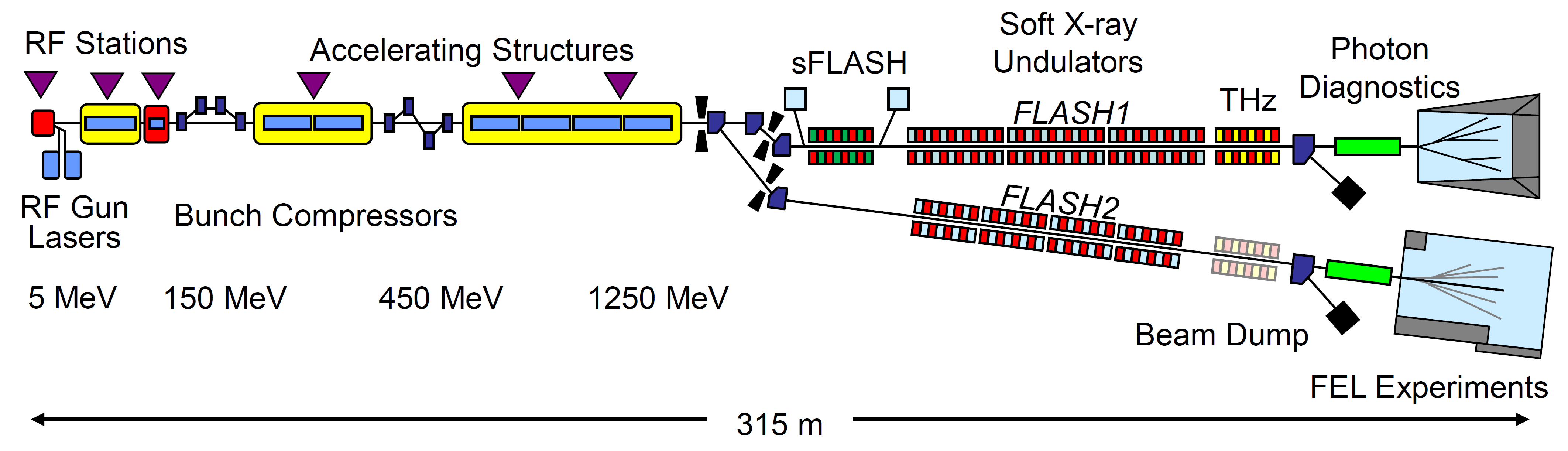}
 \caption{FLASH layout, reproduced from \cite{flash-schreiber-faatz} }
 \label{fig:flash_layout}
\end{figure}

The FLASH FEL facility at DESY, based on superconducting linac technology, first lased in 2000 at 109 nm wavelength and started user operation in 2005. It currently operates with two undulator branches 
in the wavelength range of 4 to 90 nm.
For further details see e.g. \cite{flash}, \cite{flash-schreiber-faatz}, \cite{flash12}, \cite{flash-seeding}. The facility layout is shown in Figure \ref{fig:flash_layout}. 
In this work we focused on FLASH1 since at the time of studies the diagnostics at FLASH2 was not yet fully available.
The available FLASH1 diagnostics \cite{flash-diagnostics} includes: gas monitor detectors (GMD) providing photon pulse energy and position at two locations; the MCP detectors providing photon pulse energy with less precision for a 
larger dynamic range; several photon wavelength spectrometers; electron beam position and beam loss monitors; electron bunch length diagnostics with a transverse deflecting cavity.
Due to operational constrains (the spectrometer in the tunnel blocks the GMD detectors while the beamline spectrometers require lengthy manual setup and expert operation) the spectrometers could not be effectively used
in automatic tuning. For similar reasons (difficulty with setup and interpretation of the measurement) the electron bunch length diagnostics has not been used for optimization. The electron beam position could be used, for example,
to keep the beam close to the undulator center; we however found that the beam loss minimization also accomplishes this goal.
The objective function to be maximized is thus a weighted sum of the photon pulse energy, the photon beam positions and the beam loss penalty

\begin{equation}\nonumber
 OBJ = w_{S} E_{S} - | w_{B} \cdot (r - r_0) | - LOSS
\end{equation}

\begin{figure}[h!]
 \centering
 \includegraphics[width=80mm]{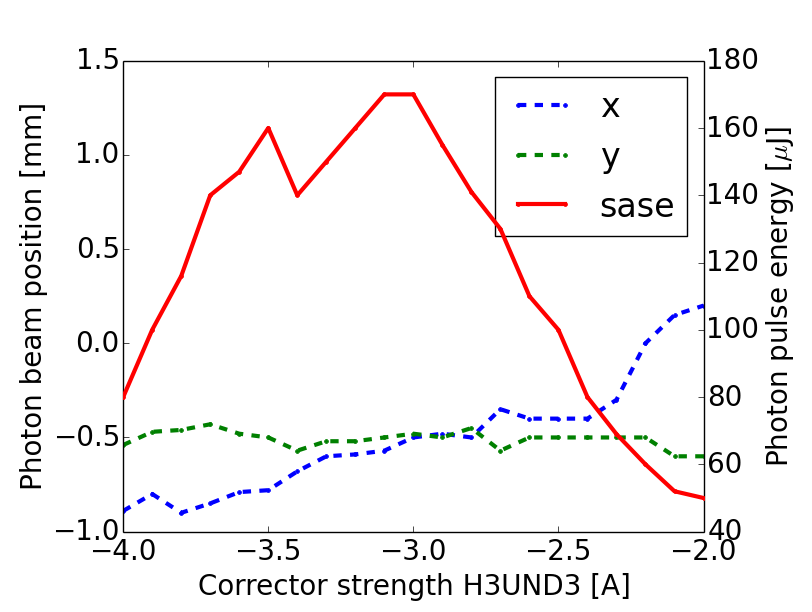}
 \caption{Photon beam position and photon pulse energy vs. H3UND4 (a horizontal orbit corrector between undulator segments) strength}
 \label{fig:photon_bpm_1}
\end{figure}

\begin{figure}[h!]
 \centering
 \includegraphics[width=80mm]{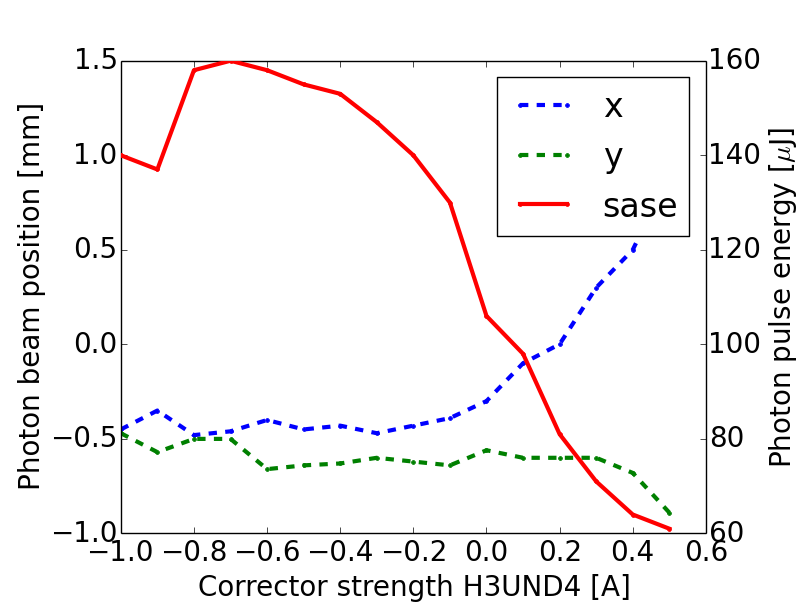}
 \caption{Photon beam position and photon pulse energy vs. H3UND3 (a horizontal orbit corrector between undulator segments) strength}
 \label{fig:photon_bpm_2}
\end{figure}

Here $LOSS$ is the beam loss penalty described previously, $E_S$ the photon pulse energy averaged over a pulse train, $w_S$ is the photon pulse energy weight, 
$r$ and $r_0$ are the measured and the target photon beam position vectors including horizontal and vertical beam positions on two monitors, $w_B$ is the corresponding weight vector, 
$|\cdot|$ could stand for either sum of absolute values or sum of squared vector components, and $\cdot$ stands for elementwise multiplication.
In practice the full objective function performs poorly and the beam pointing is usually left out. This is due to the fact that for most of the correctors, the setting for the 
optimal photon energy does not correspond to the zero photon position on the BPM. Moreover, the response of the photon BPM to a corrector setting is typically 
nonlinear. An example of such responses to some correctors between the undulator segments are shown in Figures \ref{fig:photon_bpm_1} and \ref{fig:photon_bpm_2}.
The problem of automatic optimization of photon pulse energy and pointing requires more study.
The beam loss penalty is always present, and such penalty both effectively avoids the losses and steers the electron beam close to the undulator axis.

\begin{figure}[h!]
 \centering
 \includegraphics[width=80mm]{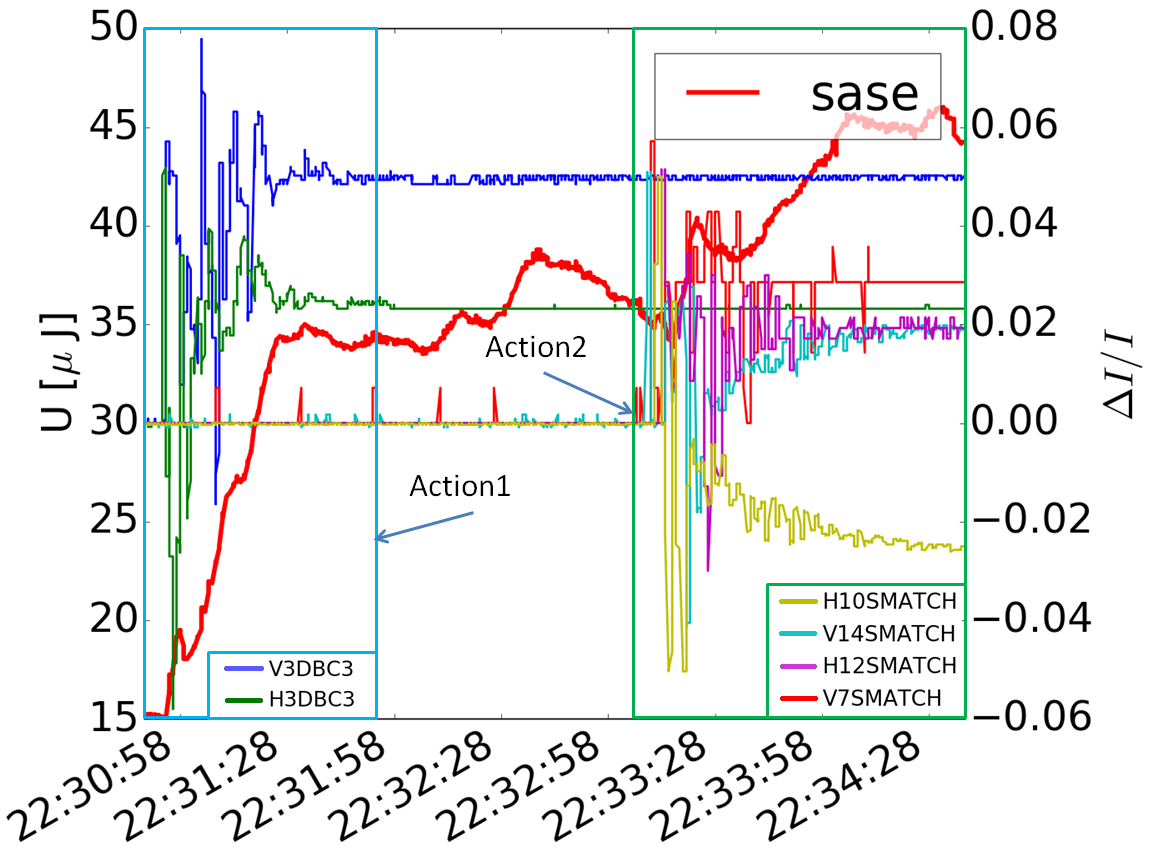}
 \caption{FLASH, an example of SASE tuning at 10.4 nm. Spikes in V7SMATCH are due to ADC resolution (device current close to zero).}
 \label{fig:flash_tuning}
\end{figure}

\begin{figure}[h!]
 \centering
 \includegraphics[width=80mm]{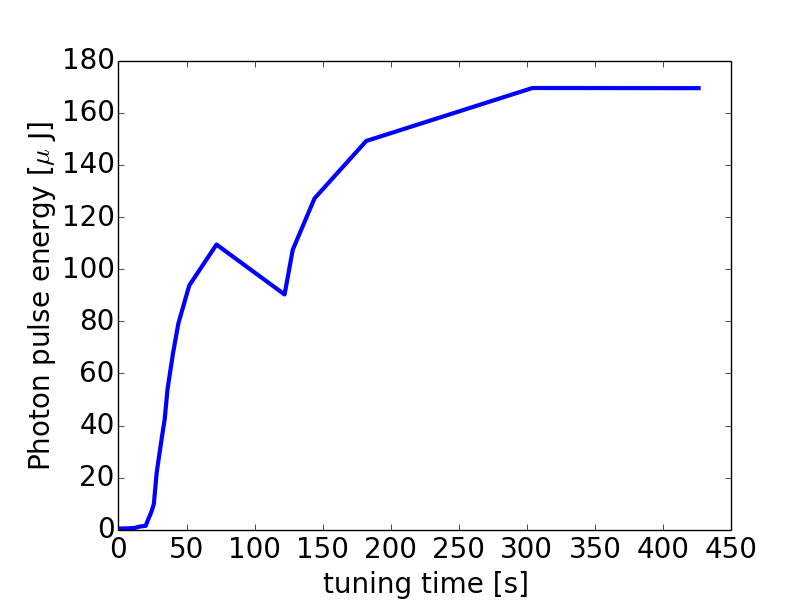}
 \caption{FLASH, an example of SASE tuning after reloading the magnet settings, setting beam transmission manually, and launching an automatic sequence 
 which included several iterations of gun solenoid,  RF phases, and launch steerer optimization, at 13.6 nm wavelength}
 \label{fig:flash_tuning2}
\end{figure}

The optimization works well in the majority of the cases, for typical photon wavelength of 10 - 20 nm and bunch charge 0.2 - 0.6 nC. Examples of optimization under two different conditions are shown in Figures 
\ref{fig:flash_tuning} and \ref{fig:flash_tuning2}.
The Nelder-Mead (simplex) and conjugate gradient (CG)  \cite{numerical_recipies} methods for the single action optimization were evaluated, 
with the latter having convergence problems due to the large uncertainty in determining the gradient of the objective function.
The simplex method is always used in practice. 2-4 devices could only be effectively used simultaneously. Two parameters could further be adjusted to improve performance: the initial optimization ``step'' (``initial simplex''), 
and the limits in which the control parameters such as magnet currents are allowed to vary. The latter is implemented by adding a penalty to the objective function. Both of them are to be set empiricaly for each device group.
If the ``initial simplex'' and the device limits are too small, the optimization may end up in a local minimum determined by short-term signal variations. On the other hand, if they are too large for certain devices such as 
strong bending magnets, the hysteresis effect could prevent the optimization from convergence, and the immediate beam loss at the first optimization step could cause the machine interlock system to switch the injector off.

The convergence of the simplex method is hard to analyze theoretically even for well-behaved functions (see e.g. \cite{simplex-convergence}), 
so it is not clear if the large number of devices leads to slow convergence speed. 
From our practical experience on the FLASH SASE FEL objective functions, the method rarely works for the number of free parameters in one optimization step greater than 6. 
An additional observation is that the magnet response time of order of a second implies that a typical machine drift
would interfere with the optimization whenever the number of objective function evaluation exceeds about 30-50 (see e.g. the photon pulse energy in Figure \ref{fig:flash_tuning} between {\it Action1} and {\it Action2} ).  
The cases where the optimization may not improve the photon pulse energy typically are

\begin{itemize}
 \item The performance has already been tuned sufficiently well
 \item There is a fluctuation in the machine which acts on a time scale of minutes, which is also a typical time scale of an optimizer run. In general, if the drifts in SASE energy are more than 20-30\% 
 on such time scale, poor tuning performance is expected (see also \cite{optim}).
 \item The control devices include a magnet with strong hysteresis 
 \item There are hardware or software problems with using the photon diagnostics or magnets. Unfortunately such situations seem to be unavoidable at present.
\end{itemize}

For example, at a typical wavelength of 13.6 nm the optimizer could often achieve the SASE pulse energy of about 160-200 $\mu J$ 
(see e.g. Figure \ref{fig:flash_tuning2} for a typical optimizer run result after loading a new ``optics file'') if both the bunch compression and the transverse optics were 
initially set to a lasing condition with SASE pulse energy of 1 $\mu J$ level. From the pulse energy data for the past two years in Figures \ref{fig:2014_sase1},\ref{fig:2015_sase1}
one sees that this is better than average performance, while still being a factor 2-3 less than the peak values achieved. The latter however required special machine setup and many hours of tuning 
and happened on relatively few occasions \cite{Schneidmiller}.

\section{Statistics for FLASH}\label{sec:statistics}

\subsection{Analysis of machine state data}\label{sec:statistics_a}

\begin{figure}[h!]
 \centering
 \includegraphics[width=80mm]{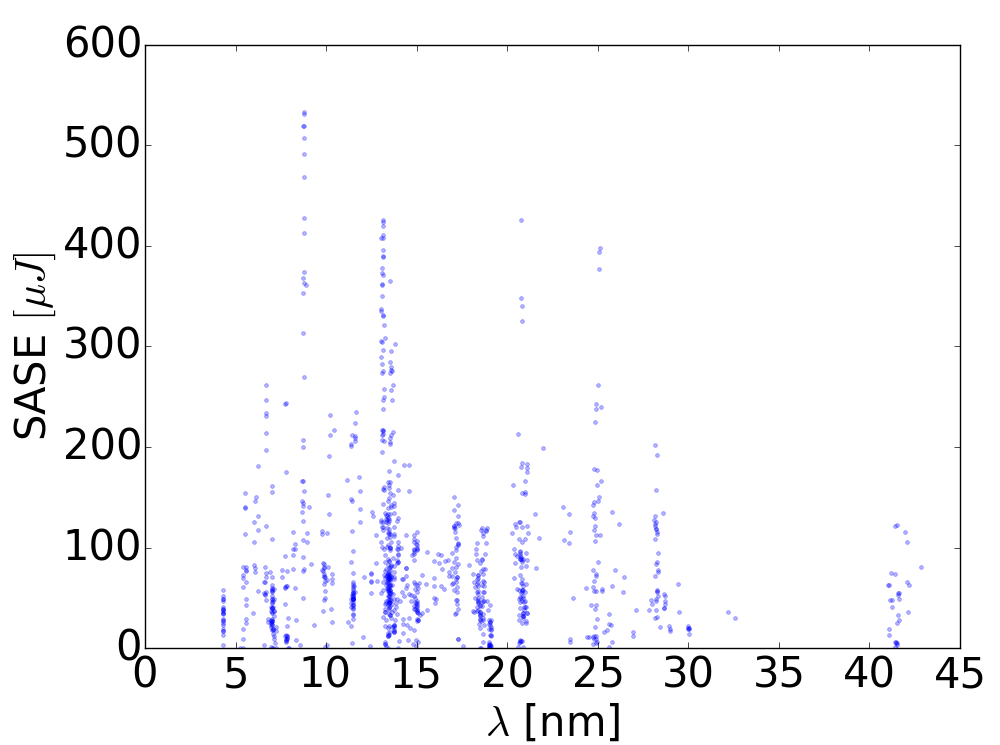}
 \caption{FLASH machine state data for 2014. Photon pulse energy vs. wavelength.}
 \label{fig:2014_sase1}
\end{figure}

\begin{figure}[h!]
 \centering
 \includegraphics[width=80mm]{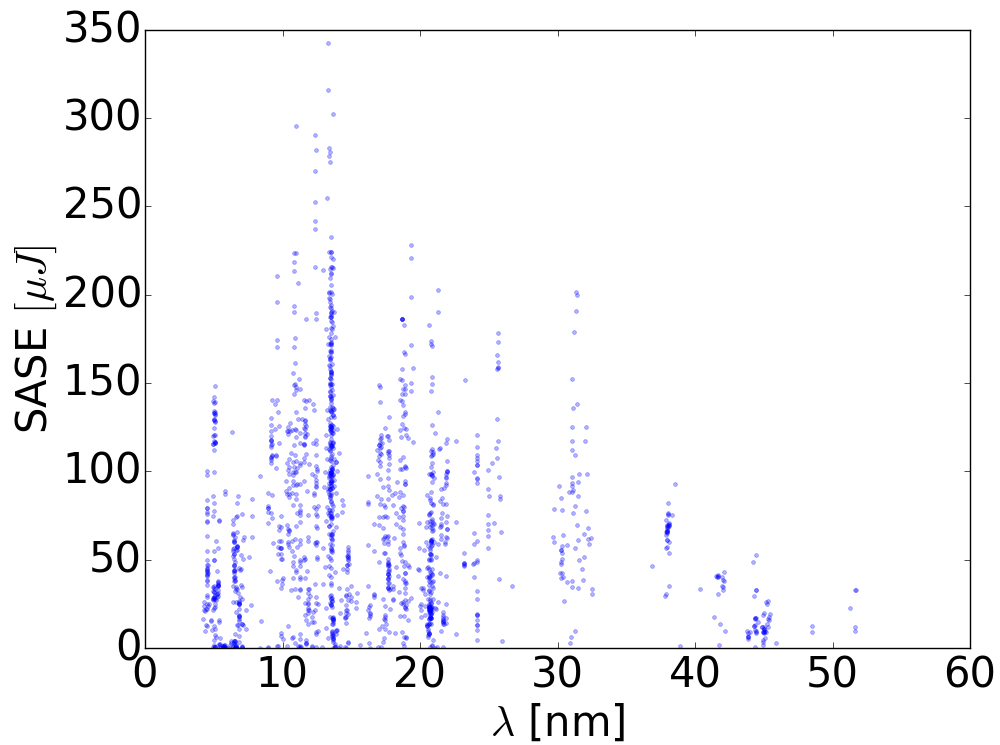}
 \caption{FLASH machine state data for 2015 and Jan-Feb 2016. Photon pulse energy vs. wavelength.}
 \label{fig:2015_sase1}
\end{figure}

\begin{figure}[h!]
 \centering
 \includegraphics[width=80mm]{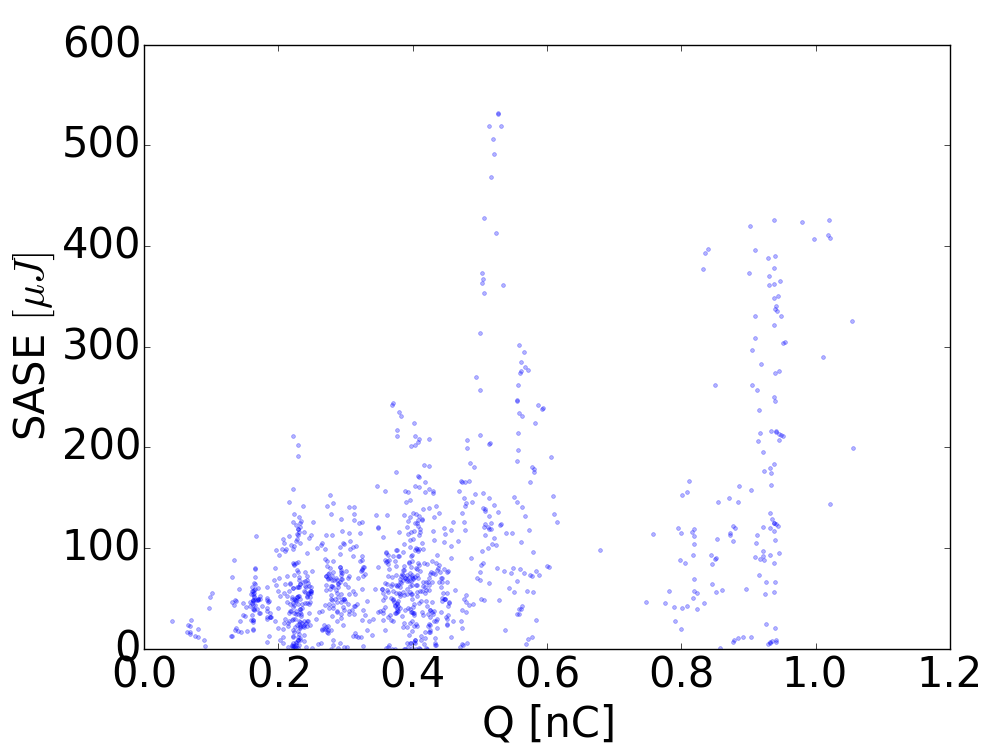}
 \caption{FLASH machine state  data for 2014. Photon pulse energy vs. bunch charge.}
 \label{fig:2014_sase2}
\end{figure}

\begin{figure}[h!]
 \centering
 \includegraphics[width=80mm]{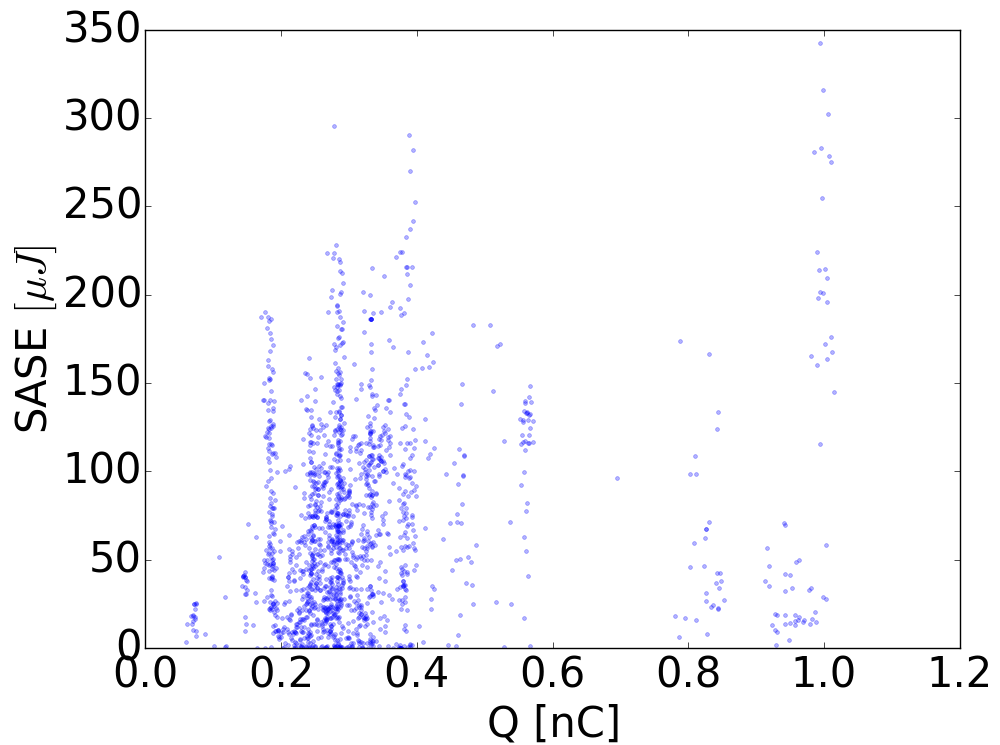}
 \caption{FLASH machine state data for 2015 and Jan-Feb 2016. Photon pulse energy vs. bunch charge.}
 \label{fig:2015_sase2}
\end{figure}

\begin{figure}[h!]
 \centering
 \includegraphics[width=80mm]{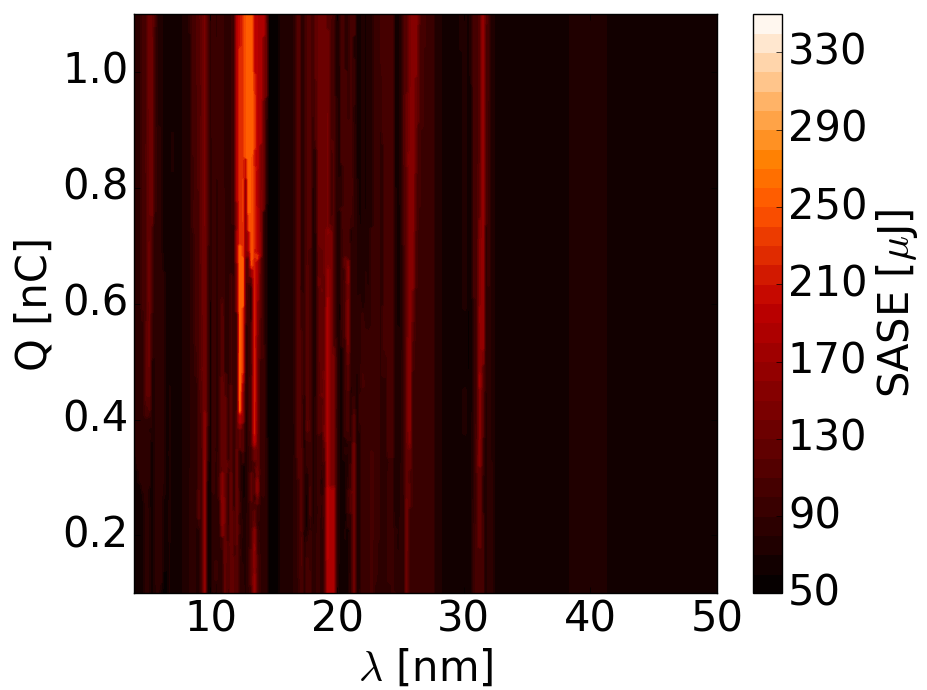}
 \caption{Expected optimal photon pulse energy as a function of wavelength and bunch charge, based on data from year 2015}
 \label{fig:perf2015}
\end{figure}

Examples of machine state data from  2014 (1063 data points) and 2015 plus 2 first months of 2016 (1645 data points) are shown in Figures \ref{fig:2014_sase1},\ref{fig:2015_sase1},\ref{fig:2014_sase2},\ref{fig:2015_sase2}.
While not immediately suggesting a tuning strategy, the machine state data can be used as a feedback for expected optimization performance, optimization stopping criteria, and evaluating if a tuning  was successful.
For example, the expected optimal photon pulse energy as a function of wavelength and bunch charge is shown in Figure \ref{fig:perf2015}.
The expected optimum here is calculated as the 5-neighbour regression \cite{stat1} with data set based on 2015 data points with pulse energy higher than 50 $\mu J$.

\subsection{Performance analysis of device groups}\label{sec:statistics_b}

The frequently used optimization knobs are listed in Tables \ref{table:perf1}-\ref{table:perf5}. The data is extracted from the tuning database (which is still relatively sparse) and the knobs ranked according to various performance criteria.
Here the devices types are as follows: H10SMATCH, H12SMATCH, V7SMATCH and V14SMATCH are the horizontal and vertical orbit correctors upstream from the undulator section (launch steerers);
H3UND1 - H3UND6 are the horizontal orbit correctors between the undulator segments; SOLENOID controls the current in the electron gun solenoid; Q3.5ECOL, Q4ECOL, Q9TCOL are quadrupoles in the dogleg section
(beam transport between linac and undulator); ACC1/SP.PHASE,  ACC23/SP.PHASE, and ACC39/SP.PHASE are the phases of the first three main accelerating modules and of the third harmonic module. 

In 55 \% of the cases the action lead to increase of the photon pulse energy, while in 20 \% of the cases the growth was more that 10 \%.
In 13 \% of the cases the photon pulse dropped significantly (more than 20 \%), which is typically a manifestation of either a hardware or low level controls software problem occurring during an optimization,
or the machine being in an unstable state with large jitters and drifts in which case the optimization procedure may not succeed.

\begin{table}[h!]
\begin{tabular}{ l | l  }
  \hline
  \bf{Devices} & \bf{Freq., \%} \\
  \hline	
  \hline
  \{H10SMATCH, H12SMATCH, V14SMATCH, & 22.9  \\
     V7SMATCH\} &    \\
  \{H3UND1,  H3UND2 ,  H3UND3, H3UND4,  & 14.6  \\
     H3UND5 , H3UND6\} &   \\
  \{SOLENOID\} &  7.3 \\
  \{H10SMATCH, V7SMATCH\} &  6.2 \\
  \{H10SMATCH, H12SMATCH\} &  6.2  \\
  \{H3UND4\} &  5.2 \\
  \hline  
\end{tabular}
\caption{Device groups sorted according to usage frequency}
\label{table:perf1}
\end{table}

\begin{table}[h!]
\begin{tabular}{ l | l  }
  \hline
  \bf{Devices} & \bf{$FoM_1$} \\
  \hline	
  \hline
  \{Q3.5ECOL, Q4ECOL, Q9TCOL\} & 74  \\
  \{H10SMATCH, H12SMATCH\} & 48  \\
  \{H10SMATCH, H12SMATCH, V14SMATCH, & 35  \\
     V7SMATCH\} &    \\
  \{ACC1/SP.PHASE, ACC23/SP.PHASE,  & 33  \\
    ACC39/SP.PHASE\} &    \\
   \{H10SMATCH, V7SMATCH\} &  27 \\
   \{H10SMATCH\} &  20 \\
  \hline  
\end{tabular}
\caption{Device groups sorted according to average $FoM_1$ (absolute pulse energy growth, $\mu J$)}
\label{table:perf2}
\end{table}

\begin{table}[h!]
\begin{tabular}{ l | l  }
  \hline
  \bf{Devices} & \bf{$FoM_1$} \\
  \hline	
  \hline
  \{H10SMATCH, H12SMATCH, V14SMATCH, & 141  \\
     V7SMATCH\} &    \\
  \{H10SMATCH, V7SMATCH\} &  94 \\
  \{H10SMATCH, H12SMATCH\} & 93  \\
  \{ACC1/SP.PHASE, ACC23/SP.PHASE,  &  84 \\
    ACC39/SP.PHASE\} &    \\
  \{Q3.5ECOL, Q4ECOL, Q9TCOL\} & 74  \\
  \{H10SMATCH\} & 33  \\
  \hline  
\end{tabular}
\caption{Device groups sorted according to maximum $FoM_1$ (absolute pulse energy growth, $\mu J$))}
\label{table:perf3}
\end{table}

\begin{table}[h!]
\begin{tabular}{ l | l  }
  \hline
  \bf{Devices} & \bf{$FoM_2$} \\
  \hline	
  \hline
    \{H10SMATCH, H12SMATCH, V14SMATCH, & 12.7  \\
     V7SMATCH\} &    \\
     \{H10SMATCH, V7SMATCH\} &  4.5 \\
     \{Q3.5ECOL, Q4ECOL, Q9TCOL\} & 0.96  \\
     \{H10SMATCH\} & 0.7  \\
     \{H10SMATCH, H12SMATCH\} & 0.7  \\
     \{ACC1/SP.PHASE, ACC23/SP.PHASE,  & 0.4  \\
    ACC39/SP.PHASE\} &    \\
  \hline  
\end{tabular}
\caption{Device groups sorted according to average $FoM_2$, (relative pulse energy growth, \%) }
\label{table:perf4}
\end{table}

\begin{table}[h!]
\begin{tabular}{ l | l  }
  \hline
  \bf{Devices} & \bf{$FoM_2$} \\
  \hline	
  \hline
  \{H10SMATCH, H12SMATCH, V14SMATCH, & 189.37  \\
     V7SMATCH\} &    \\
  \{H10SMATCH, V7SMATCH\} &  20.2 \\
  \{H3UND1,  H3UND2 ,  H3UND3, H3UND4,  & 1.4  \\
     H3UND5 , H3UND6\} &   \\
  \{H10SMATCH, H12SMATCH\} & 1.4  \\
  \{H10SMATCH\} & 1.3  \\
  \{ACC1/SP.PHASE, ACC23/SP.PHASE,  & 1.1  \\
    ACC39/SP.PHASE\} &    \\     
  \hline  
\end{tabular}
\caption{Device groups sorted according to maximum $FoM_2$, (relative pulse energy growth, \%)}
\label{table:perf5}
\end{table}

\section{Application to storage rings}\label{sec:siberia}

\begin{figure}[h!]
 \centering
 \includegraphics[width=80mm]{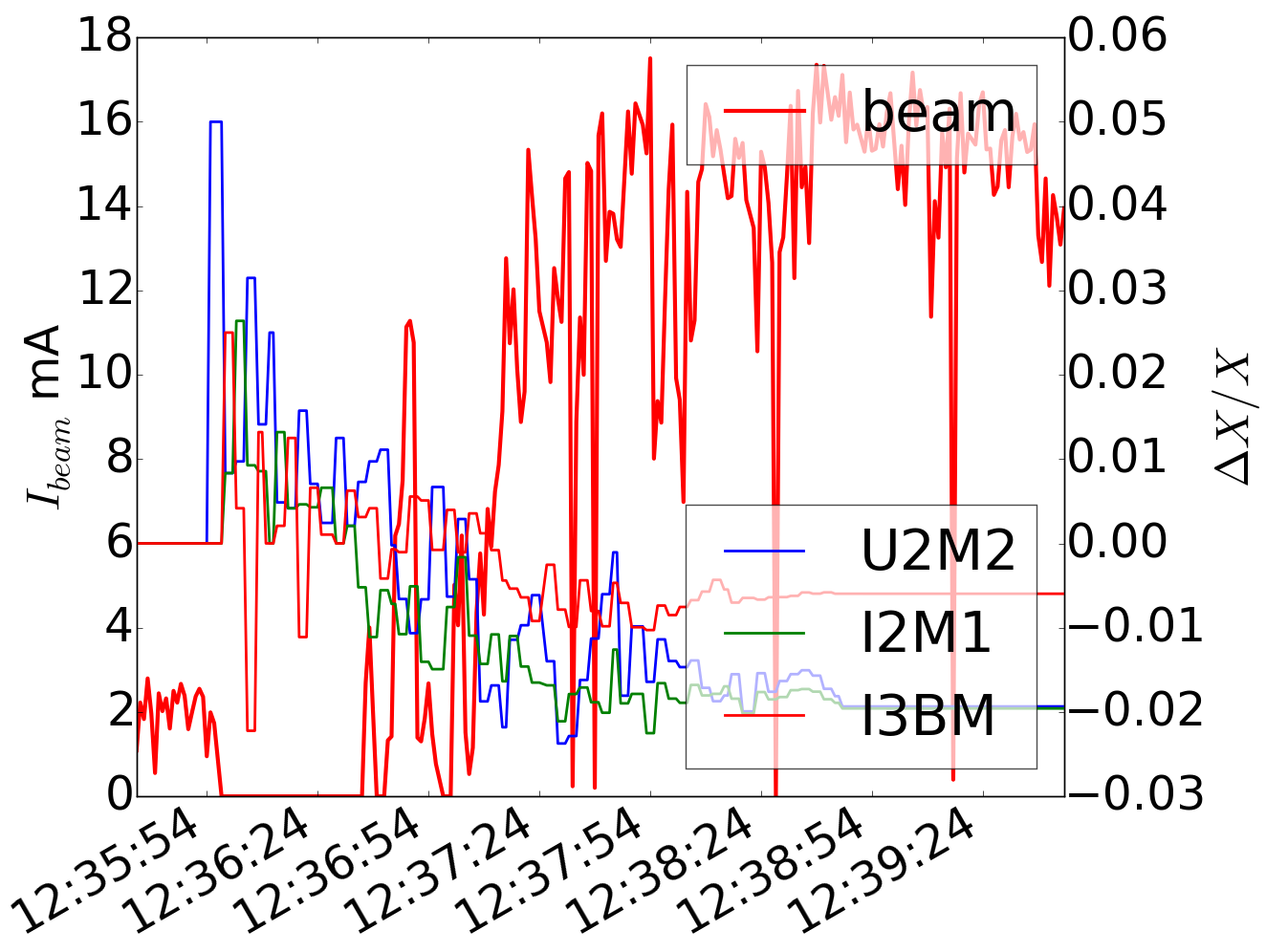}
 \caption{Injected current and the control parameter values during the optimization of injection into Siberia-2 }
 \label{fig:siberia}
\end{figure}

Since both the software and the approach are completely general, they can be applied to a wide variety of optimization problems. 
So, the software was adapted to optimize the injection efficiency into the Siberia-2 storage ring (for more details see \cite{optim2}).
The most effective tuning parameters there are the dipole magnet strength in the transport channel from the linac to the Siberia-1 booster (I2M1), 
the septum voltage (U2M2) and the main dipole strength in the Siberia-2 ring (I3BM). 
The objective is the current stored in the main ring after a single injection and subsequent dump.
An example of optimization is shown in Figure \ref{fig:siberia}. The apparent drops in the injected current are due to no bunches coming from the gun, which
is in turn related to timing problems. The procedure is however sufficiently robust and not affected by such missing bunches. The injection efficiency is tuned to an optimum level 
(similar to what an experienced operator typically achieves) in a matter of minutes.

\section{Conclusion}

In this paper we presented a simple and robust empirical tuning proceeder, which shows good performance for SASE FEL optimization as well as for optimization of other accelerator performance parameters, 
for linacs and for storage rings.
While the hardware performance ultimately limits the performance of the optimizer, many tuning procedures can be potentially automated. 
The performance achieved on routine tasks is often comparable to that of an experience operator, which could save considerable amount of time in future.

Machine learning potentially offers a fruitful direction in improving the empirical optimization methods by analysing machine state and optimizer performance data.
We have presented some examples of such analysis including knob ranking and orbit data clustering.
Including more diagnostics data in such analysis and extending the methods are subjects for future studies.

\section{Acknowledgements} The authors wish to acknowledge the support of many people who contributed considerably to our studies, in particular V. Ayvazyan,
 C. Behrens, R. Brinkmann,  W. Decking, B. Faatz, L. Fr\"ohlich,  K. Honkavaara, R. Kammering, V. Kocharyan, T. Limberg,  E. Shneydmiller, S. Schreiber, and M. Vogt. 
The  work was partially supported by Ioffe Roentgen Institute grant EDYN EMRAD.


\begin{thebibliography}{99}

\bibitem{flash} W. Ackermann et al., Nature Photonics,  2007, 1, 331

\bibitem{kek} J.W. Flanagan et al., ``A simple real-time beam tuning program for the KEKB injector linac", KEK, Tsukuba, Japan, KEK Preprint 98-209 (1999)
\bibitem{huang} X. Huang, J. Corbett, J. Safranek, J. Wu, ``An algorithm for online optimization of accelerators", Nucl.Instrum.Meth. A726 (2013) 77-83 (2013)
\bibitem{fermi} G. Gaio, M. Lonza, ``Automatic FEL optimization at FERMI", Proceedings of ICALEPCS2015, Melbourne, Australia (2015)


\bibitem{ocelot} I. Agapov et al., ``OCELOT: a software framework for synchrotron light source and FEL studies", Nucl. Instr. Meth. A. 768 (2014) pp. 151-156
\bibitem{dohlus} M. Dohlus, ``Interpretation of FLASH measurements on June 16th 2011", http://www.desy.de/fel-beam/data/talks/files/FLASH16thJune.pdf
 \bibitem{scholz} M. Scholz, J. Zemella, M. Vogt, ``Beam Optics Measurements at FLASH2", in Proceedings of IPAC 2015, Richmond, USA (2015)
\bibitem{optim2} S. Tomin, I. Agapov, G. Geloni, I. Zagorodnov et al., ``Progress in Automatic Software-based Optimization of Accelerator Performance", in Proceedings of IPAC 2016, Busan, Korea (2016)
\bibitem{optim3} M. McIntire, T. Cope, D. Ratner, S. Ermon, ``Bayesian Optimization of FEL Performance at LCLS", in Proceedings of IPAC 2016, Busan, Korea (2016)

\bibitem{xfeltdr} M. Altarelli et al. (Eds.), ``The European X-Ray Free Electron Laser, Technical Design Report", DESY 2006-097, 2006.

\bibitem{optim} I. Agapov, G. Geloni, I. Zagorodnov,  ``Statistical optimization of FEL performance", in Proceedings of IPAC 2015, Richmond, USA (2015)

\bibitem{ocelot2} I. Agapov, M. Dohlus, G. Geloni, S. Tomin, I. Zagorodnov, ``FEL Simulations with OCELOT", in Proceedings of IPAC 2015, Richmond, USA (2015);
                  I. Zagorodnov, I. Agapov, M. Dohlus, S. Tomin, submitted to PAC 2016

\bibitem{stat1} T. Hastie, R. Tibshirani, J. Friedman, ``The elements of statistical learning; data mining, inference, and prediction", Springer series in statistics, 2001

\bibitem{doocs} http://doocs.desy.de/
\bibitem{sklearn} http://scikit-learn.org/stable/
\bibitem{scipy} https://www.scipy.org/
\bibitem{flash-schreiber-faatz} S. Schreiber and B. Faatz, High Power Laser Science and Engineering, 2015, 3, e20 doi:10.1017/hpl.2015.16; https://flash.desy.de/
\bibitem{flash12} M. Scholz, B. Faatz, S. Schreiber et al, ``First simultaneous operation of two SASE beamlines in FLASH", in Proceedings of FEL 2015, Daejon, Korea (2015)

\bibitem{flash-seeding} J. Boedewadt, S. Ackermann, R. Assmann et al., ``Recent results from FEL seeding at FLASH" , in Proceedings of IPAC 2015, Richmond, USA (2015)
\bibitem{flash-diagnostics} K. Tiedtke et al., ``The soft x-ray free-electron laser FLASH at DESY: beamlines, diagnostics and end-stations", New Journal of Physics  11 (2009)

\bibitem{numerical_recipies} W. Press, S. Teukolsky, W. Vetterling, B. Flannery,  ``Numerical recipies in C++", 2nd Edition, Cambridge University Press (2002)
\bibitem{simplex-convergence}  J. Lagarias, J. Reeds, M. Wright, P. Wright, ``Convergence properties of the Nelder-Mead simplex method in low dimensions", SIAM J. Optim, Vol 9. No. 1, pp. 112-147 (1998)
\bibitem{Schneidmiller} E. Shneydmiller, private communications.



\end{thebibliography}
\end{document}